\documentclass[
  ,draft            
  ]
  {aipproc}

\layoutstyle{6x9}

\begin{document}

\title{The First Sources of Light}

\author{Volker Bromm}{address={Astronomy Department, Harvard
  University, 60 Garden St., Cambridge, MA 02138} }

\author{Abraham Loeb}{ address={Astronomy Department, Harvard
  University, 60 Garden St., Cambridge, MA 02138}
  ,altaddress={Institute for Advanced Study, Princeton, NJ 08540;
  Guggenheim Fellow} }

\begin{abstract}
We review recent theoretical results on the formation of the first
stars and quasars in the universe, and emphasize related open
questions. In particular, we list important differences between the
star formation process at high redshifts and in the present-day
universe. We address the importance of heavy elements in bringing
about the transition from an early star formation mode dominated by
massive stars, to the familiar mode dominated by low mass stars, at
later times. We show how gamma-ray bursts can be utilized to probe the
first epoch of star formation.  Finally, we discuss how the first
supermassive black holes could have formed through the direct collapse
of primordial gas clouds.
\end{abstract}

\date{}
\maketitle


\section{Introduction}

The first sources of light ionized \cite{WyL03a,Cen03} and metal-enriched
\cite{FL03} the intergalactic medium (IGM) and consequently had
important effects on subsequent galaxy formation \cite{BL2001} and on
the large-scale polarization anisotropies of the cosmic microwave
background \cite{Kap02}.  {\it When did the cosmic dark ages end?}  In
the context of popular cold dark matter (CDM) models of hierarchical
structure formation, the first stars are predicted to have formed in
dark matter halos of mass $\sim 10^{6}M_{\odot}$ that collapsed at
redshifts $z\simeq 20-30$ \cite{BL2001,Yos2003}. The first quasars, on
the other hand, are likely to have formed in more massive host
systems, at redshifts $z \ge 10$ \cite{HL2001}, and certainly before
$z\sim 6.4$, the redshift of the most distant quasar known
\cite{Fan2003}.

Results from recent numerical simulations of the collapse and
fragmentation of primordial clouds suggest that the first stars were
predominantly very massive, with typical masses $M_{\ast}\ge 100
M_{\odot}$ \cite{BCL1999,BCL2002,NaU2001,ABN2002}.  Despite the
progress already made, many important questions remain unanswered; the
purpose of this brief review is to discuss these open questions and to
put them in perspective. An example for an open question is: {\it How
does the primordial initial mass function (IMF) look like?}  Having
constrained the characteristic mass scale, still leaves undetermined
the overall range of stellar masses and the power-law slope which is
likely to be a function of mass. In addition, it is presently unknown
whether binaries or, more generally, clusters of zero-metallicity
stars, can form.  Evidently, the observational signature as well as
the fate of the first stars depend sensitively on whether primordial
star formation is predominantly clustered or isolated. This in turn is
affected by the nature of the feedback that the first stars exert on
their surroundings.  The first stars are expected to produce copious
amounts of UV photons and to possibly explode as energetic
hypernovae. {\it How effective will their negative feedback be in
suppressing star formation in neighboring high-density clumps?}

Predicting the properties of the first stars is important for the
design of upcoming instruments, such as the {\it James Webb Space
Telescope} (JWST) \footnote{See http:// ngst.gsfc.nasa.gov.}, or the
next generation of large ($>10$m) ground-based telescopes.  The hope
is that over the upcoming decade, it will become possible to confront
current theoretical predictions about the properties of the first
sources of light with direct observational data. The increasing volume
of new data on high redshift galaxies and quasars from existing
ground-based telescopes, signals the emergence of this new frontier in
cosmology.

\section{Star Formation Then and Now}

Currently, we do not have direct observational constraints on how
the first stars, the so-called Population~III stars, formed at
the end of the cosmic dark ages. It is, therefore, instructive to
briefly summarize what we have learned about star formation in the
present-day universe, where theoretical reasoning is guided by a
wealth of observational data (see \cite{Pud2002} for a recent review).

Population~I stars form out of cold, dense molecular gas that is
structured in a complex, highly inhomogeneous way. The molecular
clouds are supported against gravity by turbulent velocity fields and
pervaded on large scales by magnetic fields.  Stars tend to form in
clusters, ranging from a few hundred up to $\sim 10^{6}$ stars. It
appears likely that the clustered nature of star formation leads to
complicated dynamics and tidal interactions that transport angular
momentum, thus allowing the collapsing gas to overcome the classical
centrifugal barrier \cite{Lar2002}.  The IMF of Pop~I stars is
observed to have the approximate Salpeter form (e.g., \cite{Kr02})
\begin{equation}
\frac{{\rm d}N}{{\rm d log}M}\propto M^{x} \mbox{\ ,}
\end{equation}
where
\begin{equation}
x\simeq \left\{
\begin{array}{rl}
-1.35 & \mbox{for \ }M\ge 0.5 M_{\odot}\\
0.0 & \mbox{for \ }0.007 \le M\le 0.5 M_{\odot}\\
\end{array}
\right. \mbox{\ .}
\end{equation}
The lower cutoff in mass corresponds roughly to the opacity limit for
fragmentation. This limit reflects the minimum fragment mass, set when
the rate at which gravitational energy is released during the collapse
exceeds the rate at which the gas can cool (e.g., \cite{MJR1976}).
The most important feature of the observed IMF is that $\sim 1
M_{\odot}$ is the characteristic mass scale of Pop~I star formation,
in the sense that most of the mass goes into stars with masses close
to this value. In Figure 1, we show the result from a recent hydrodynamical
simulation of the collapse and fragmentation of a molecular cloud core
\cite{BBB2002,BBB2003}.  This simulation illustrates the highly
dynamic and chaotic nature of the star formation process\footnote{See
http:// www.ukaff.ac.uk/starcluster for an animation.}.

\begin{figure}
  \includegraphics[height=.3\textheight]{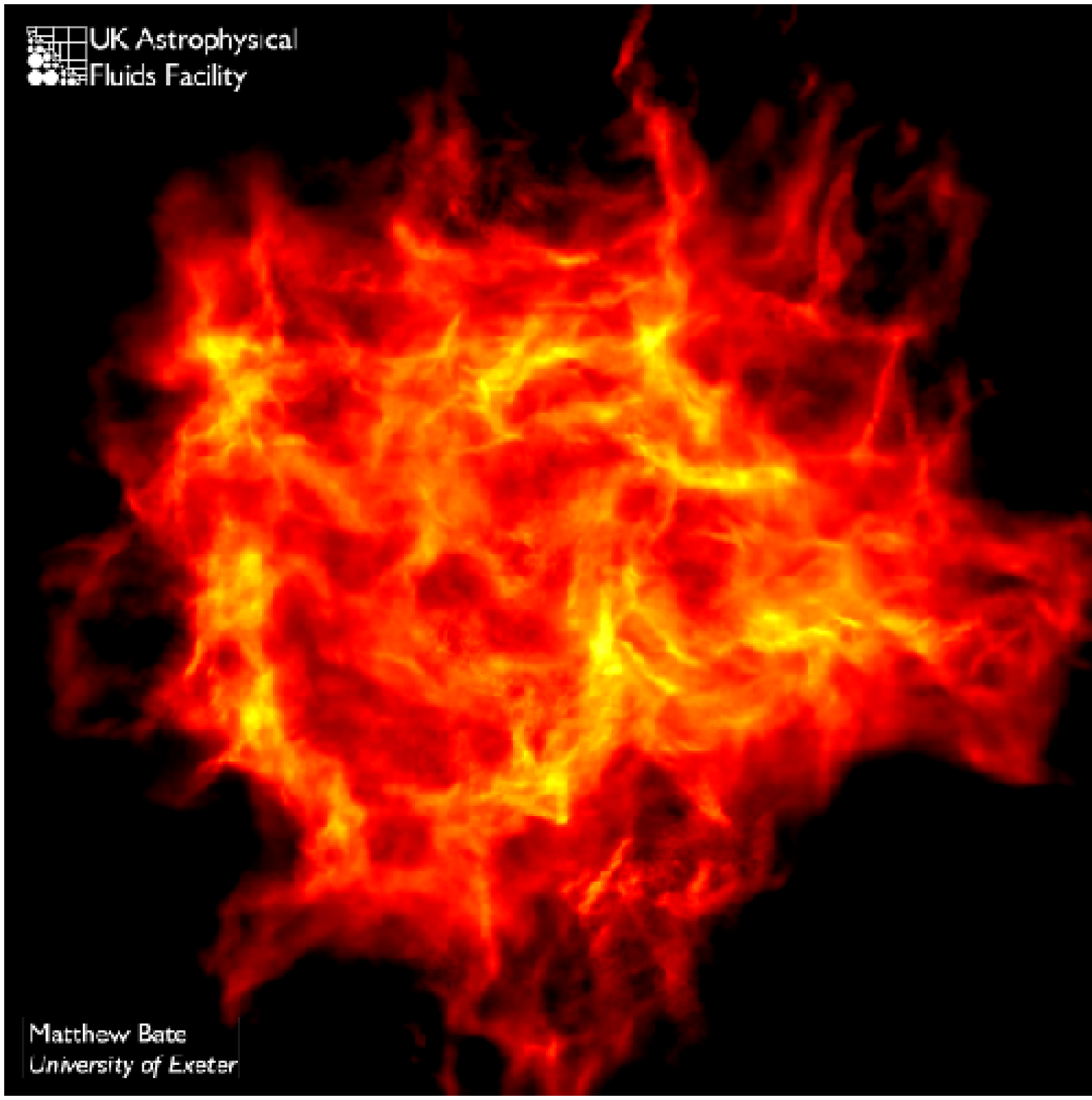}
  \includegraphics[height=.3\textheight]{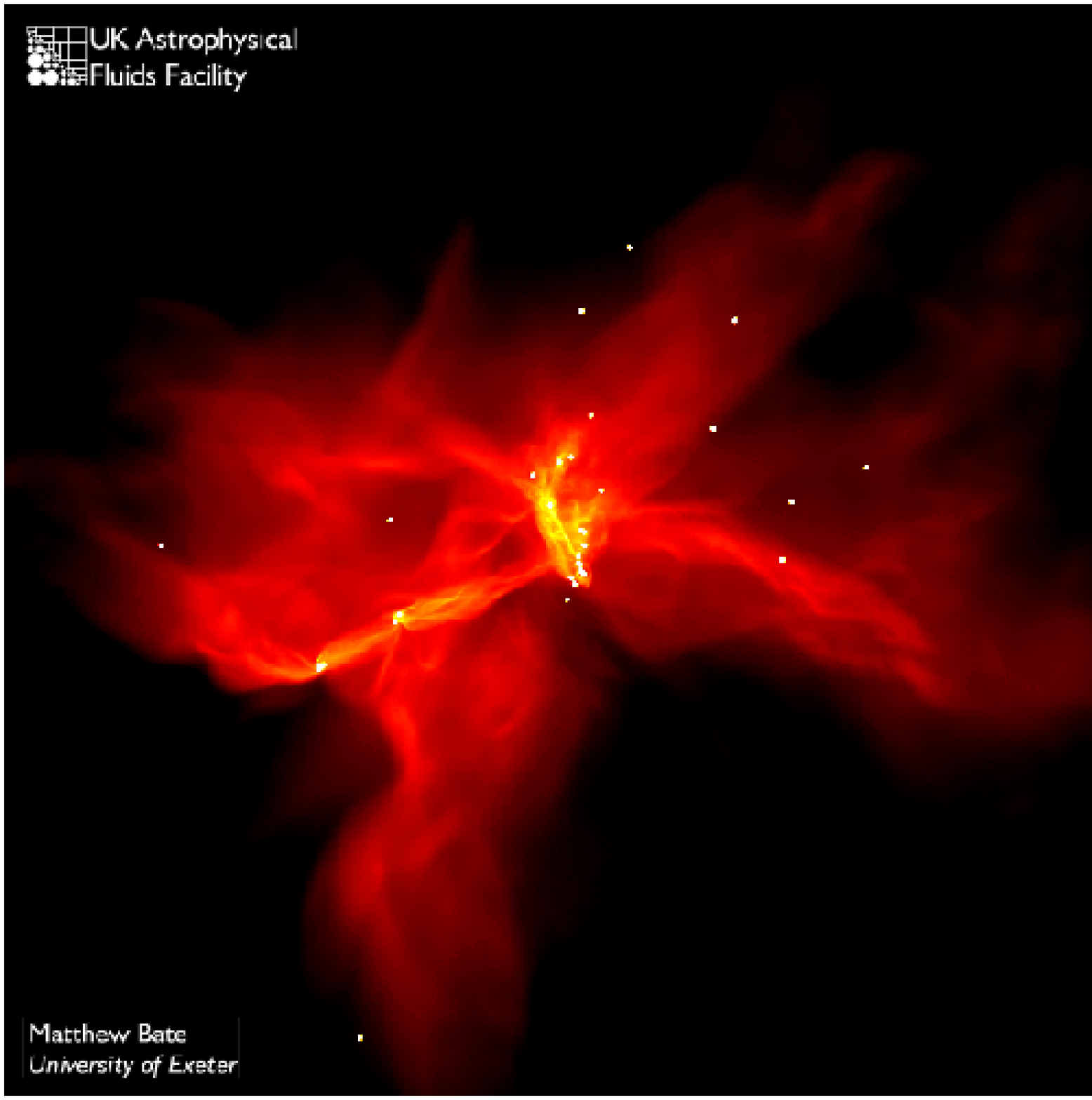}
\caption{A hydrodynamic simulation of the collapse and fragmentation
of a turbulent molecular cloud in the present-day universe (from
\cite{BBB2003}).  The cloud has a mass of $50 M_{\odot}$.  The panels
show the column density through the cloud, and span a scale of 0.4 pc
across.  {\it Left:} The initial phase of the collapse. The turbulence
organizes the gas into a network of filaments, and decays thereafter
through shocks.  {\it Right:} A snapshot taken near the end of the
simulation, after 1.4 initial free-fall times of $2\times 10^{5}$yr.
Fragmentation has resulted in $\sim 50$ stars and brown dwarfs.
The star formation efficiency is $\sim 10$\% on the scale of the
overall cloud, but can be much larger in the dense sub-condensations.
This result is in good agreement with what is observed in local star-forming
regions.}
\end{figure}

The metal-rich chemistry, magnetohydrodynamics, and radiative
transfer involved in present-day star formation is complex, and we
still lack a comprehensive theoretical framework that predicts the IMF
from first principles. Star formation in the high redshift universe,
on the other hand, poses a theoretically more tractable problem due to
a number of simplifying features, such as: (i) the initial absence of
heavy metals and therefore of dust; and (ii) the absence of
dynamically-significant magnetic fields, in the pristine gas left over
from the big bang. The cooling of the primordial gas does then only
depend on hydrogen in its atomic and molecular form.  Whereas in the
present-day interstellar medium, the initial state of the star forming
cloud is poorly constrained, the corresponding initial conditions for
primordial star formation are simple, given by the popular
$\Lambda$CDM model of cosmological structure formation. We now turn to
a discussion of this theoretically attractive and important problem.

\section{Primordial Star Formation}

{\it How did the first stars form?} A complete answer to this question
would entail a theoretical prediction for the Population~III IMF,
which is rather challenging. Let us start by addressing the simpler
problem of estimating the characteristic mass scale of the first
stars. As mentioned before, this mass scale is observed to be $\sim 1
M_{\odot}$ in the present-day universe.  To investigate the collapse
and fragmentation of primordial gas, we have carried out numerical
simulations, using the smoothed particle hydrodynamics (SPH)
method. We have included the chemistry and cooling physics relevant
for the evolution of metal-free gas (see \cite{BCL2002} for
details). Improving on earlier work \cite{BCL1999,BCL2002} by
initializing our simulation according to the $\Lambda$CDM model, we
focus here on an isolated overdense region that corresponds to a
3$\sigma-$peak: a halo containing a total mass of $10^{6}M_{\odot}$,
and collapsing at a redshift $z_{\rm vir}\simeq 20$.

In Figure~2 ({\it left panel}), we show the gas density within the
central $\sim 25$~pc, briefly after the first high-density clump has
formed as a result of gravitational runaway collapse. Once the gas has
exceeded a threshold density of $10^{7}$ cm$^{-3}$, a sink particle is
inserted into the simulation to replace it.  
This choice for the density threshold ensures that the local Jeans mass is
resolved throughout the simulation.
The clump (i.e., sink particle) has an initial mass of $M_{\rm Cl}\sim
10^{3}M_{\odot}$, and grows subsequently by ongoing accretion of
surrounding gas.  High-density clumps with such masses result from the
chemistry and cooling rate of molecular hydrogen, H$_{2}$, which
imprint characteristic values of temperature, $T\sim 200$~K, and
density, $n\sim 10^{4}$ cm$^{-3}$, into the metal-free gas (see
\cite{BCL2002}).  Evaluating the Jeans mass for these characteristic
values results in $M_{J}\sim 10^{3}M_{\odot}$, which is close to the
initial clump masses found in the simulations.

\begin{figure}
  \includegraphics[height=.3\textheight]{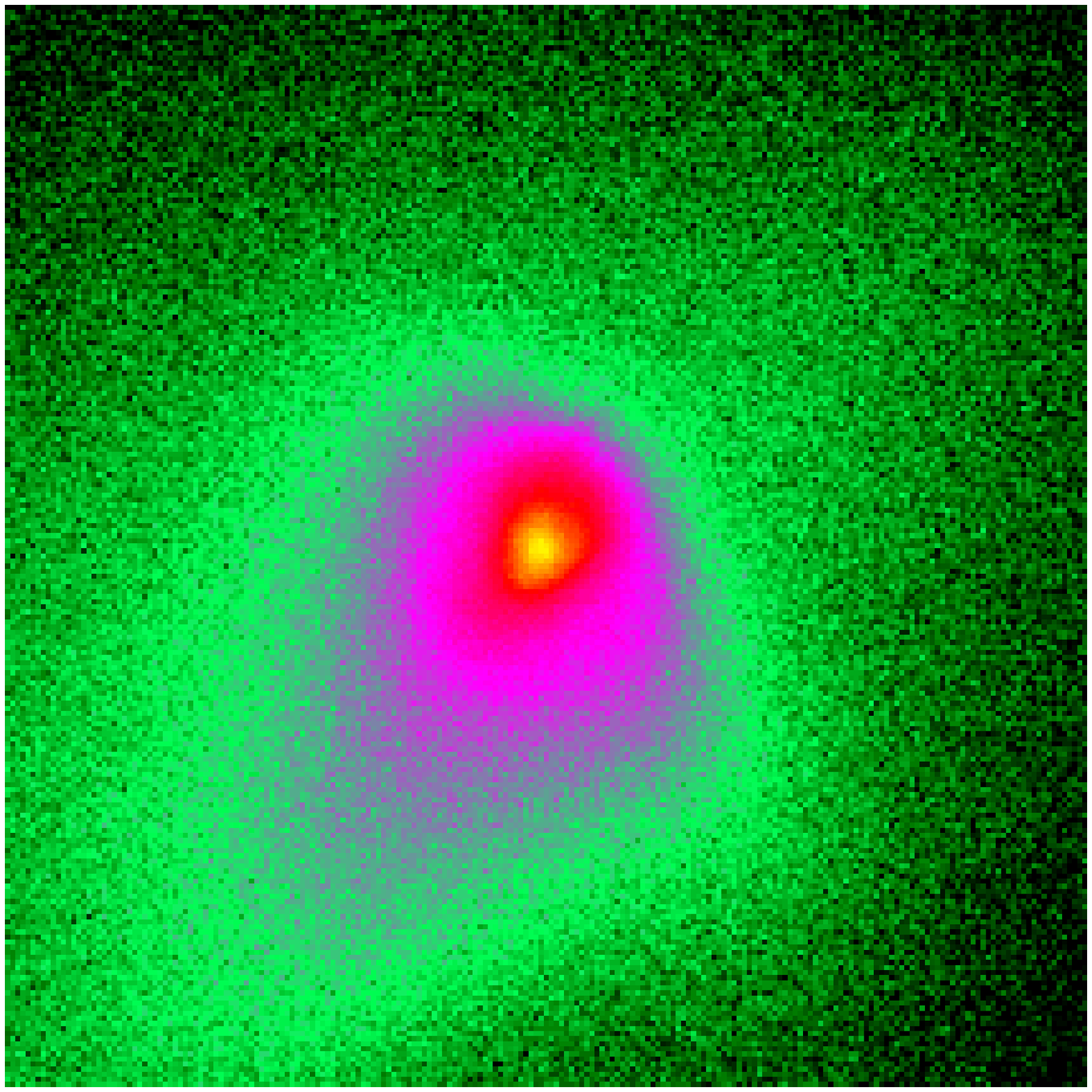}
  \includegraphics[height=.3\textheight]{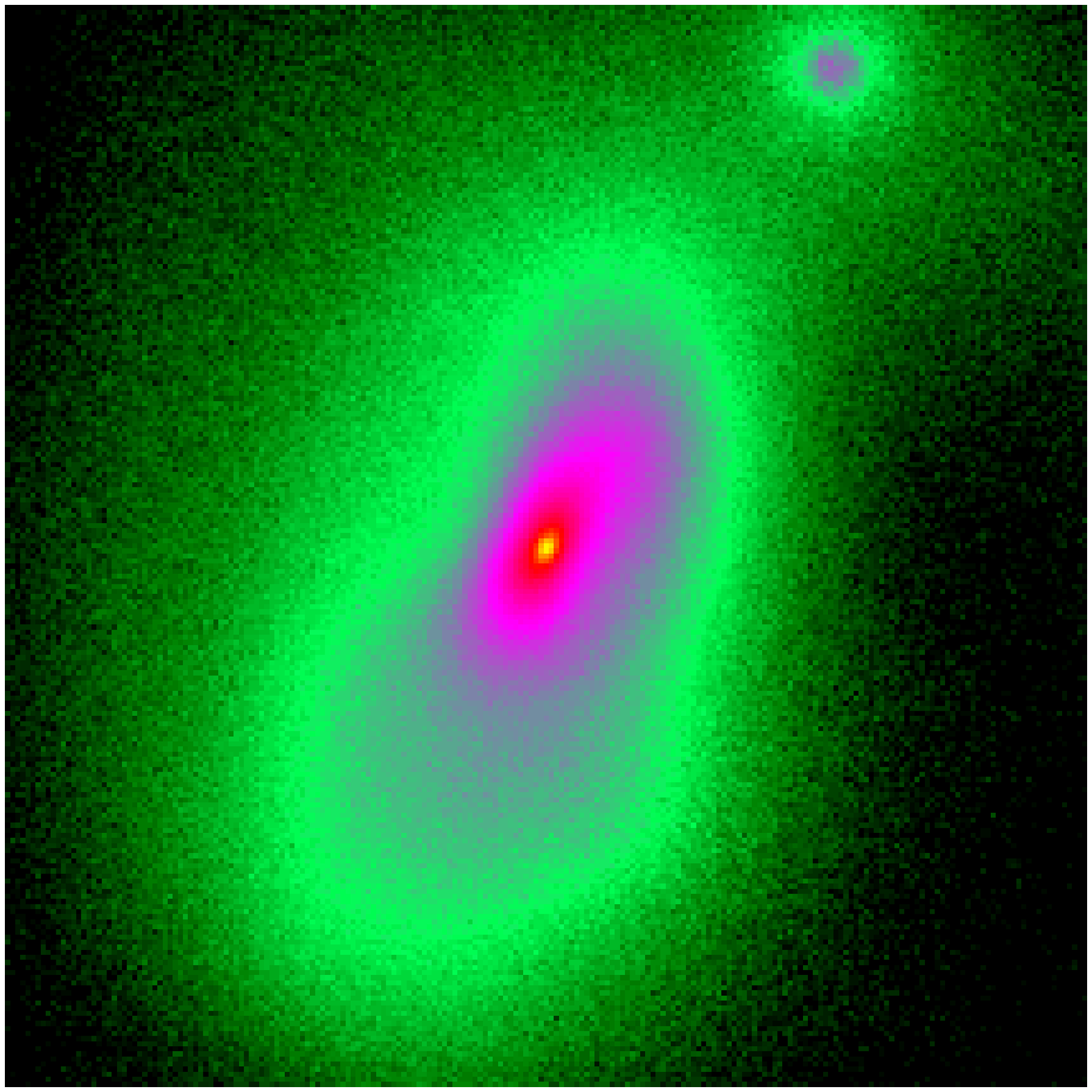}
\caption{Collapse and fragmentation of a primordial cloud.  Shown is
the projected gas density at a redshift $z\simeq 21.5$, briefly after
gravitational runaway collapse has commenced in the center of the
cloud.  {\it Left:} The coarse-grained morphology in a box with linear
physical size of 23.5~pc.  
At this time in the unrefined simulation, a high-density clump (sink
particle) has formed with an initial mass of $\sim 10^{3}M_{\odot}$.
{\it Right:} The fine-grain morphology in a box with linear physical
size of 0.5~pc.
The central density peak, vigorously gaining mass by accretion, is
accompanied by a secondary clump.}
\end{figure}

\begin{figure}[ht]
  \includegraphics[height=.3\textheight]{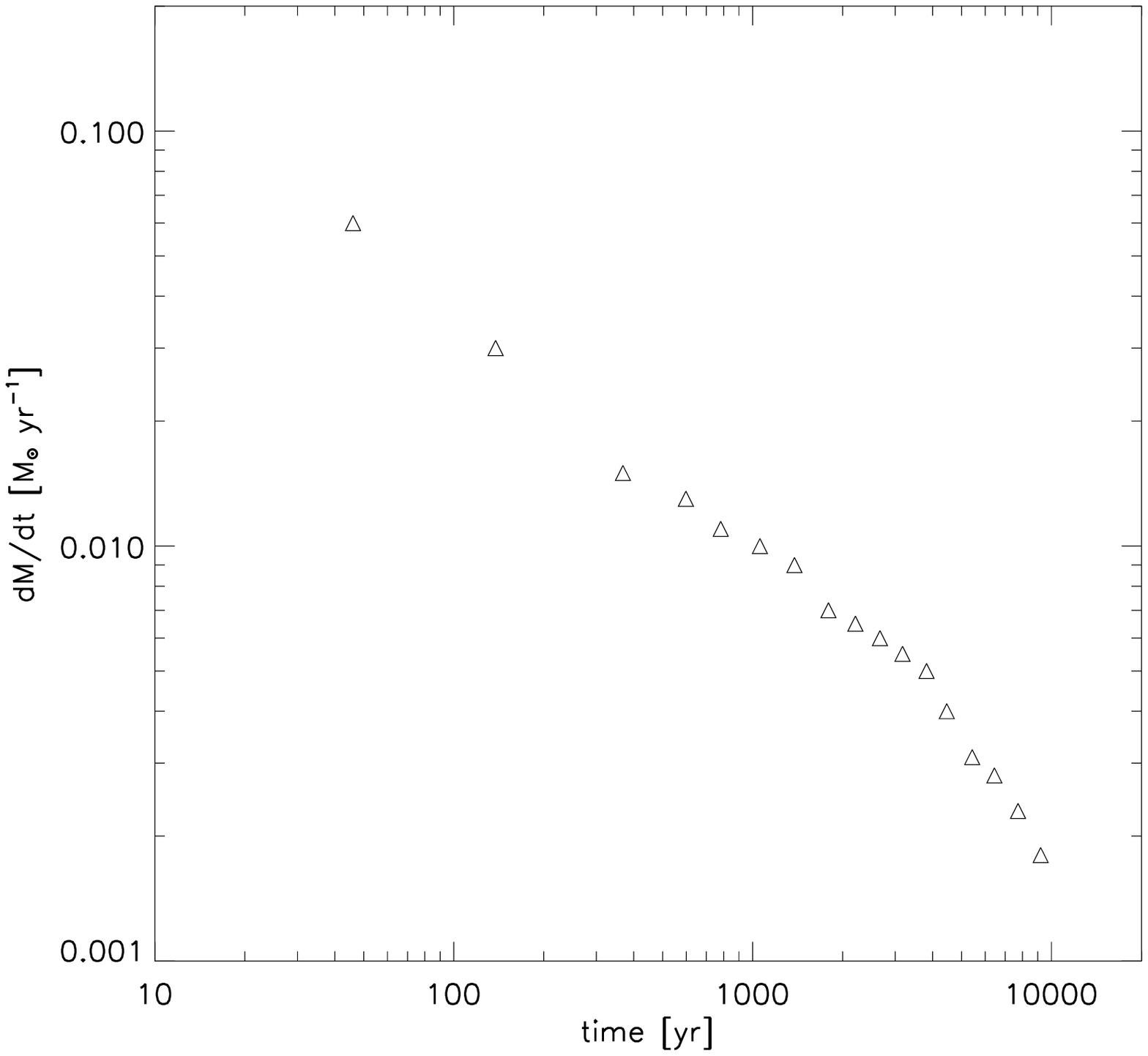}
  \includegraphics[height=.3\textheight]{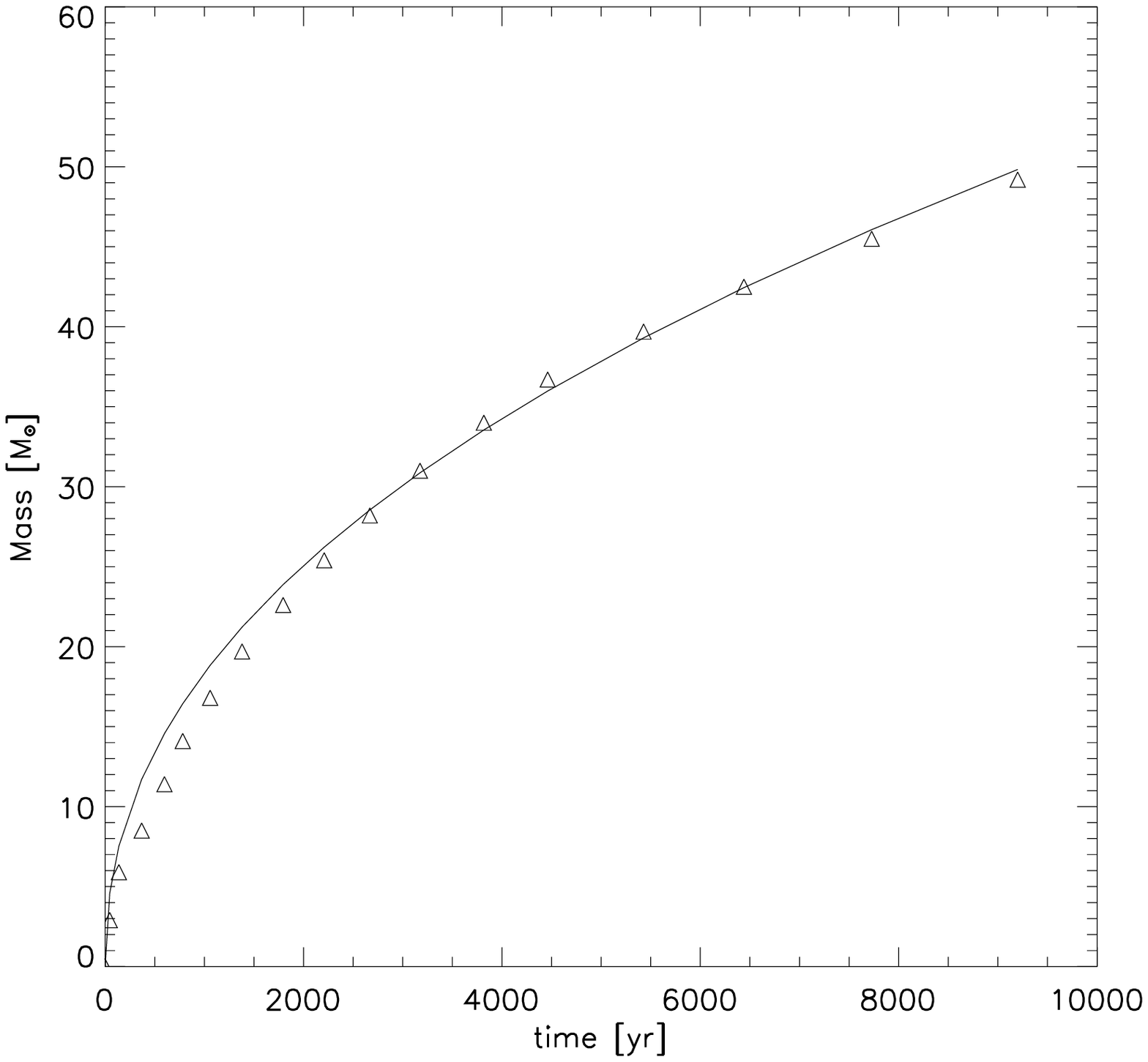}
\caption{Accretion onto a primordial protostar.  The morphology of
this accretion flow is shown in Fig.~2.  {\it Left:} Accretion rate
(in $M_{\odot}$~yr$^{-1}$) vs. time (in yr) since molecular core
formation.  {\it Right:} Mass of the central core (in $M_{\odot}$)
vs. time.  {\it Solid line:} Accretion history approximated as:
$M_{\ast}\propto t^{0.45}$.  Using this analytical approximation, we
extrapolate that the protostellar mass has grown to $\sim 150
M_{\odot}$ after $\sim 10^{5}$~yr, and to $\sim 700 M_{\odot}$ after
$\sim 3\times 10^{6}$~yr, the total lifetime of a very massive star.  }
\end{figure}

The high-density clumps are clearly not stars yet. To probe the
subsequent fate of a clump, we have re-simulated the evolution of the
central clump with sufficient resolution to follow the collapse to
higher densities (see \cite{BrL2003} for a description of the
refinement technique). In Figure~2 ({\it right panel}), we show the
gas density on a scale of 0.5~pc, which is two orders of magnitude
smaller than before. Several features are evident in this plot. First,
the central clump does not undergo further sub-fragmentation, and is
likely to form a single Population~III star. Second, a companion clump
is visible at a distance of $\sim 0.25$~pc. If negative feedback from
the first-forming star is ignored, this companion clump would undergo
runaway collapse on its own approximately $\sim 3$~Myr later.  This
timescale is comparable to the lifetime of a very massive star
(VMS)\cite{BKL2001}.  If the second clump was able to survive the
intense radiative heating from its neighbor, it could become a star
before the first one explodes as a supernova (SN). Whether more than
one star can form in a low-mass halo thus crucially depends on the
degree of synchronization of clump formation.  Finally, the
non-axisymmetric disturbance induced by the companion clump, as well
as the angular momentum stored in the orbital motion of the binary
system, allow the system to overcome the angular momentum barrier for
the collapse of the central clump (see \cite{Lar2002}).

The recent discovery of the star HE0107-5240 with a mass of $0.8 M_{\odot}$
and an iron abundance of ${\rm [Fe/H]} = -5.3$ \cite{Cr02} shows that at least
some low mass stars could have formed out of extremely low-metallicity gas.
Our simulations show that although the majority of clumps are very massive,
a few of them, like the secondary clump in Fig.~2, are significantly less
massive. Alternatively, low-mass fragments could form in the dense, 
shock-compressed shells that surround the first hypernovae \cite{MBH03}.

{\it How massive were the first stars?} Star formation typically
proceeds from the `inside-out', through the accretion of gas onto a
central hydrostatic core.  Whereas the initial mass of the hydrostatic
core is very similar for primordial and present-day star formation
\cite{ON1998}, the accretion process -- ultimately responsible for
setting the final stellar mass, is expected to be rather different. On
dimensional grounds, the accretion rate is simply related to the sound
speed cubed over Newton's constant (or equivalently given by the ratio
of the Jeans mass and the free-fall time): $\dot{M}_{\rm acc}\sim
c_s^3/G \propto T^{3/2}$. A simple comparison of the temperatures in
present-day star forming regions ($T\sim 10$~K) with those in
primordial ones ($T\sim 200-300$~K) already indicates a difference in
the accretion rate of more than two orders of magnitude.

Our refined simulation enables us to study the three-dimensional
accretion flow around the protostar (see also
\cite{OP2001,Rip2002,Tan2003}).  We now allow the gas to reach
densities of $10^{12}$ cm$^{-3}$ before being incorporated into a
central sink particle. At these high densities, three-body reactions
\cite{PSS1983} have converted the gas into a fully molecular form.  In
Figure~3, we show how the molecular core grows in mass over the first
$\sim 10^{4}$~yr after its formation. The accretion rate ({\it left
panel}) is initially very high, $\dot{M}_{\rm acc}\sim 0.1
M_{\odot}$~yr$^{-1}$, and subsequently declines according to a power
law, with a possible break at $\sim 5000$~yr. The mass of the
molecular core ({\it right panel}), taken as an estimator of the
proto-stellar mass, grows approximately as: $M_{\ast}\sim \int
\dot{M}_{\rm acc}{\rm d}t \propto t^{0.45}$. A rough upper limit for
the final mass of the star is then: $M_{\ast}(t=3\times 10^{6}{\rm
yr})\sim 700 M_{\odot}$. In deriving this upper bound, we have
conservatively assumed that accretion cannot go on for longer than the
total lifetime of a VMS.

{\it Can a Population~III star ever reach this asymptotic mass limit?}
The answer to this question is not yet known with any certainty, and
it depends on whether the accretion from a dust-free envelope is
eventually terminated by feedback from the star (e.g.,
\cite{OP2001,Rip2002,Tan2003,OI2002}). The standard mechanism by which
accretion may be terminated in metal-rich gas, namely radiation
pressure on dust grains \cite{WC1987}, is evidently not effective for
gas with a primordial composition. Recently, it has been speculated
that accretion could instead be turned off through the formation of an
H~II region \cite{OI2002}, or through the radiation pressure exerted
by trapped Ly$\alpha$ photons \cite{Tan2003}. The termination of the
accretion process defines the current unsolved frontier in studies of
Population~III star formation. Current simulations indicate that the
first stars were predominantly very massive, and consequently rather
different from present-day stellar populations. The crucial question
then arises: {\it How and when did the transition take place from the
early formation of massive stars to that of low-mass stars at later
times?}  We address this problem next.

\section{The Second Generation of Stars}

The very first stars, marking the cosmic Renaissance of structure
formation, formed under conditions that were much simpler than the
highly complex environment in present-day molecular clouds.
Subsequently, however, the situation rapidly became more complicated
again due to the feedback from the first stars on the IGM.  Supernova
explosions dispersed the nucleosynthetic products from the first
generation of stars into the surrounding gas (e.g., \cite{MFR01,MFM02,
TSD02}), including also dust grains produced in the explosion itself
\cite{LH97,TodF01}.  Atomic and molecular cooling became much more
efficient after the addition of these metals. Moreover, the
presence of ionizing cosmic rays, as well as of UV and X-ray
background photons, modified the thermal and chemical behavior of
the gas in important ways (e.g., \cite{MBA01,MBA03}).

Early metal enrichment was likely the dominant effect that brought
about the transition from Population~III to Population~II star
formation.  Recent numerical simulations of collapsing primordial
objects with overall masses of $\sim 10^{6}M_{\odot}$, have shown that
the gas has to be enriched with heavy elements to a minimum level of
$Z_{\rm crit}\simeq 10^{-3.5}Z_{\odot}$, in order to have any effect
on the dynamics and fragmentation properties of the system
\cite{Om00,BFCL01}.  Normal, low-mass (Population~II) stars are
hypothesized to only form out of gas with metallicity $Z\ge Z_{\rm
crit}$.  Thus, the characteristic mass scale for star formation is
expected to be a function of metallicity, with a discontinuity at
$Z_{\rm crit}$ where the mass scale changes by $\sim$ two orders of
magnitude. The redshift where this transition occurs has important
implications for the early growth of cosmic structure, and the
resulting observational signature (e.g.,
\cite{WyL03a,FL03,MBH03,Sch02}).

Important caveats, however, remain. The determination of the critical
metallicity mentioned above \cite{BFCL01} implicitly assumes that the
gas at temperatures below $\sim 8000$~K is maintained in ionization
equilibrium by cosmic rays, with an ionization rate that is scaled
from the Galactic value by the factor $Z/Z_{\odot}$. The cosmic-ray
flux in the early universe might well have not obeyed this simple
relation, and it is not clear whether cosmic rays could have
successfully `activated' the metals.

\section{Gamma-ray Bursts as Probes of the First Stars}

Gamma-ray bursts (GRBs) are the brightest electromagnetic explosions
in the universe, and they should be detectable out to redshifts $z>10$
\cite{LR00,Cia00}. Although the nature of the central engine that
powers the relativistic jets is still debated, recent evidence
indicates that GRBs trace the formation of massive stars
\cite{Bloom01,kul,Tot97,Wij98,BlNat00}.  Since the first stars are
predicted to be predominantly very massive, their death might possibly
give rise to GRBs at very high redshifts.  A detection of the
highest-redshift GRBs would probe the earliest epochs of star
formation, one massive star at a time. The upcoming {\it Swift}
satellite\footnote{See http://swift.gsfc.nasa.gov.}, planned for
launch in late 2003, is expected to detect about a hundred GRBs per
year. The redshifts of high-$z$ GRBs can be easily measured through
infrared photometry, based on the Gunn-Peterson trough in their spectra due to
Ly$\alpha$ absorption by neutral intergalactic hydrogen along the line of sight.
{\it Which fraction of the detected bursts will originate at
redshifts $z\ge 5$?}

To assess the utility of GRBs as probes of the first stars, we have
calculated the expected redshift distribution of GRBs \cite{BrL2002}.
Under the assumption that the GRB rate is simply proportional to the
star formation rate, we find that about a quarter of all GRBs detected
by {\it Swift}, will originate from a redshift $z\ge 5$ (see
Fig.~4). This estimate is rather uncertain because of the poorly
determined GRB luminosity function.  We caution that the rate of
high-redshift GRBs may be significantly suppressed if the early
massive stars fail to launch a relativistic outflow. This is
conceivable, as metal-free stars may experience negligible mass loss
before exploding as a supernova.  They would then retain their massive
hydrogen envelope, and any relativistic jet might be quenched before
escaping from the star \cite{Heg03}.

\begin{figure}
  \includegraphics[height=.5\textheight,width=.8\textwidth]{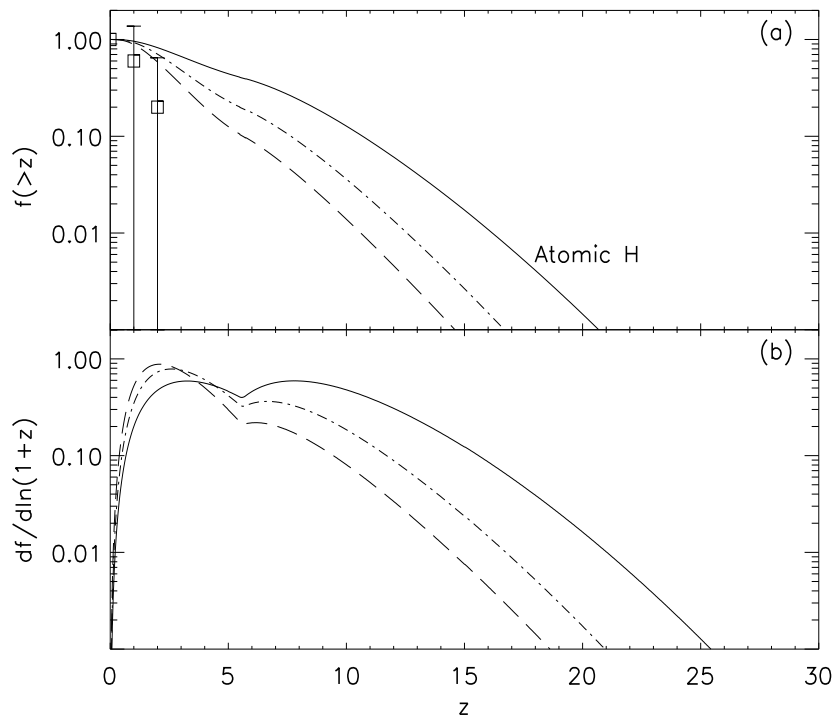}
\caption{Redshift distribution of all GRBs in comparison with
measurements in flux-limited surveys (from \cite{BrL2002}).  ({\it a})
Fraction of bursts that originate at a redshift higher than $z$
vs. $z$.  The data points reflect $\sim 20$ observed redshifts to
date.  ({\it b}) Fraction of bursts per logarithmic interval of
$(1+z)$ vs. $z$.  {\it Solid lines:} All GRBs for star formation in
halos massive enough to allow cooling via lines of atomic hydrogen.
The calculation assumes that the GRB rate is proportional (with a
constant factor) to the star formation rate at all redshifts.  {\it
Dot-dashed lines:} Expected distribution for {\it Swift}.  {\it
Long-dashed lines:} Expected distribution for BATSE.  The curves for
the two flux-limited surveys are rather uncertain because of the
poorly-determined GRB luminosity function.  }
\end{figure}

If high-redshift GRBs exist, the launch of {\it Swift} later
this year will open up an exciting new window into the cosmic dark
ages.  In difference from quasars or galaxies that fade with
increasing redshift, GRB afterglows maintain a roughly constant
observed flux at different redshifts for a fixed observed time lag
after the $\gamma$-ray trigger \cite{Cia00}. The increase in the
luminosity distance at higher redshifts is compensated by the fact
that a fixed observed time lag corresponds to an intrinsic time
shorter by a factor of $(1+z)$ in the source rest-frame, during which
the GRB afterglow emission is brighter. This quality makes GRB
afterglows the best probes of the metallicity and ionization state of
the intervening IGM during the epoch of reionization. In difference
from quasars, the UV emission from GRBs has a negligible effect on the
surrounding IGM (since $\sim 10^{51}$ ergs can only ionize $\sim 4\times
10^4M_\odot$ of hydrogen). Moreover, the host galaxies of GRBs induce a much
weaker perturbation to the Hubble flow in the surrounding IGM, compared to
the massive hosts of the brightest quasars \cite{BL03}. Hence,
GRB afterglows offer the ideal probe (much better than quasars or
bright galaxies) of the damping wing of the Gunn-Peterson trough
\cite{Mi98} that signals the neutral fraction of the IGM as a function
of redshift during the epoch of reionization.

\begin{figure}
  \includegraphics[height=.3\textheight]{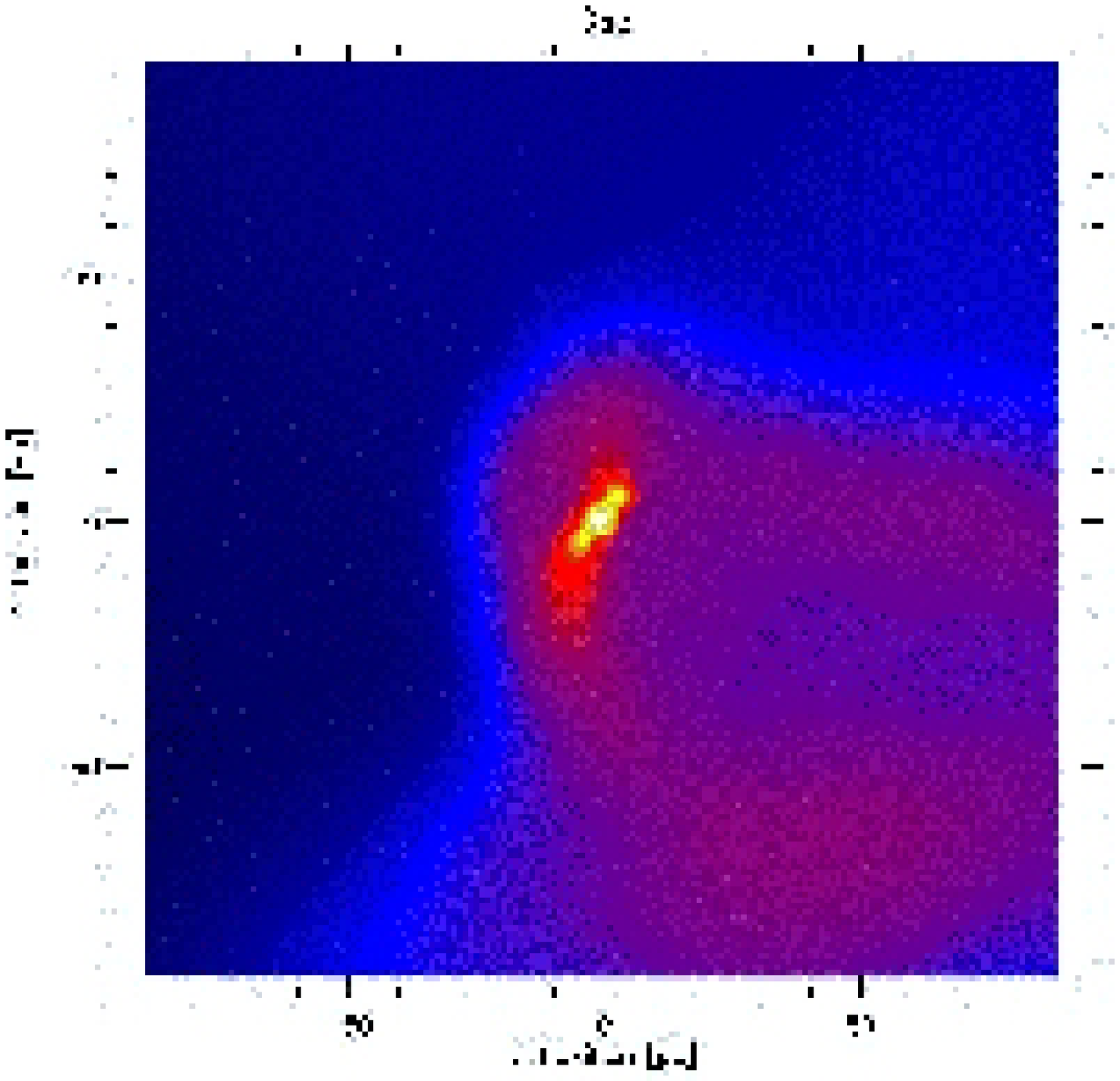}
  \includegraphics[height=.3\textheight]{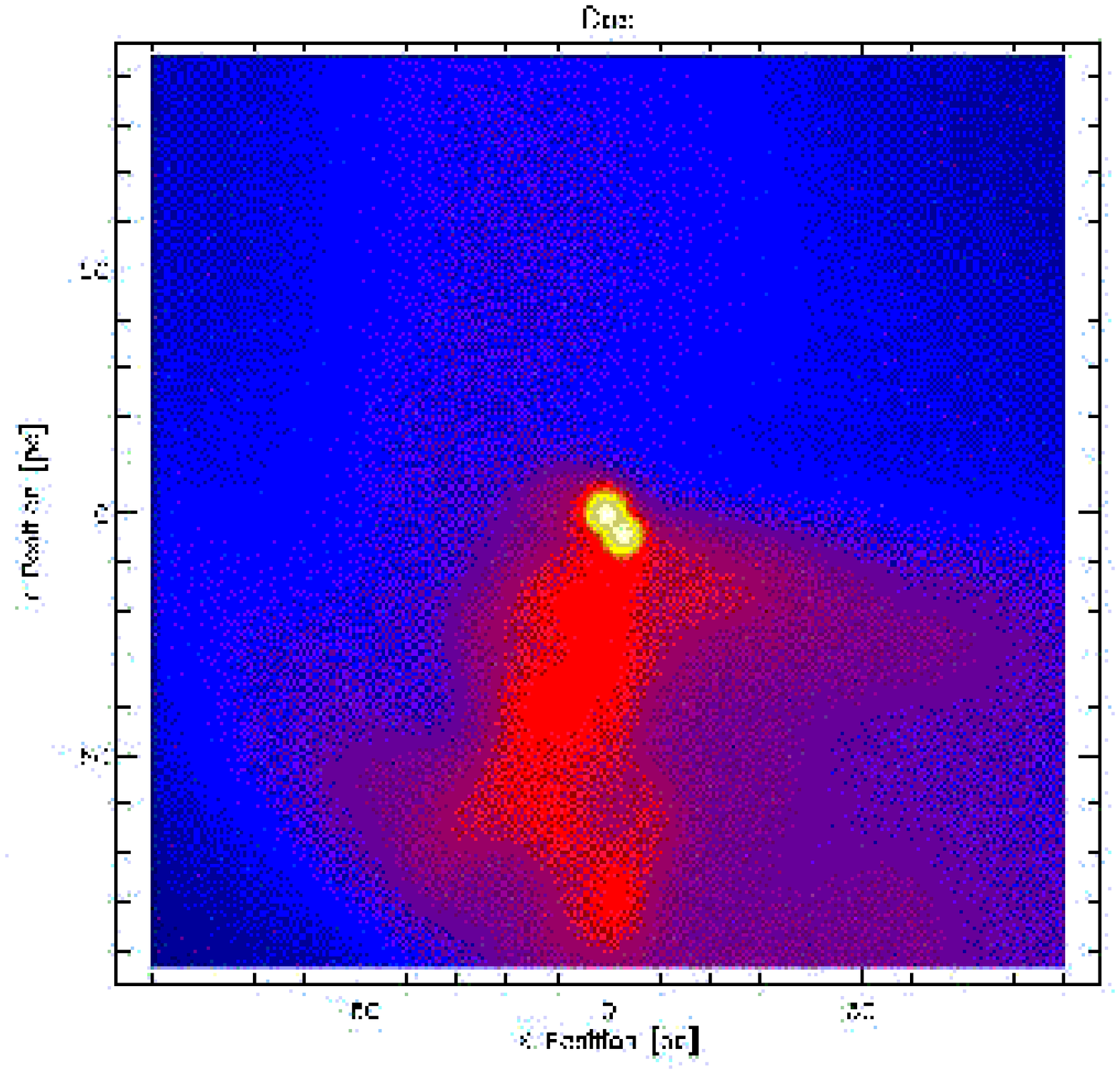}
\caption{Central gas densities in simulations of two dwarf galaxies
just above the atomic cooling threshold at $z\sim 10$ with different
initial spins and no H$_2$ molecules (from \cite{BrL2003}).  Shown is
the projection in the $x-y$ plane, and the box size is 200 pc on a
side.  {\it Left:} Case with zero initial spin. One compact object has
formed in the center with a mass of $2.7 \times 10^{6}M_{\odot}$ and a
radius of $\le 1$~pc.  {\it Right:} Case with an initial spin of
$\lambda=0.05$. Here, two compact objects have formed with masses
of $2.2 \times 10^{6}M_{\odot}$ and $3.1 \times 10^{6}M_{\odot}$,
respectively, and radii $\le 1$~pc.}
\end{figure}

\section{The First Quasars}

Quasars are believed to be powered by the heat generated during the
accretion of gas onto supermassive black holes (SMBHs; see e.g.,
\cite{Rees1984} ).  The existence of SMBHs with inferred masses of
$\ge 10^{9}M_{\odot}$, less than a billion years after the big bang,
as implied by the recent discovery of quasars at redshifs $z\ge 6$
\cite{Fan2003,Beck2001,Djor2001,Fan2002}, provides important
constraints on the SMBH formation scenario \cite{HL2001}.  The
brightest quasars at $z\ge 6$ are most likely hosted by rare galaxies,
more massive than $\sim 10^{12}M_\odot$ \cite{BL03}, that end up today as
the most massive elliptical galaxies \cite{LP02}.  {\it Can the seeds
of SMBHs form through the direct collapse of primordial gas clouds at
high redshifts?} Previous work \cite{LR1994} has shown that without a
pre-existing central point mass, this is rendered difficult by the
negative feedback resulting from star formation in the collapsing
cloud. The input of kinetic energy due to supernova explosions
prevents the gas from assembling in the center of the dark matter
potential well, thus precluding the direct formation of a SMBH. If,
however, star formation was suppressed in a cloud that could still
undergo overall collapse, such an adverse feedback would not occur.

We have carried out SPH simulations \cite{BrL2003} of isolated
2$\sigma-$peaks with total masses of $10^{8}M_{\odot}$ that collapse
at $z_{\rm vir}\sim 10$. The virial temperature of these dwarf
galaxies exceeds $\sim 10^4$K, so as to allow collapse of their gas
through cooling by atomic hydrogen transitions \cite{OH02}. Since
structure formation proceeds in a bottom-up fashion, such a system
would encompass lower-mass halos that would have collapsed earlier
on. These sub-systems have virial temperatures below $10^{4}$~K, and
consequently rely on the presence of H$_{2}$ for their
cooling. Molecular hydrogen, however, is fragile and readily destroyed
by photons in the Lyman-Werner bands with energies
($11.2-13.6$~eV) just below the Lyman limit \cite{HRL1997}.  These
photons are able to penetrate a predominantly neutral IGM.

We first consider the limiting case in which H$_{2}$ destruction is
complete.  Depending on the initial spin, which is a measure of the
degree of rotational support in the cloud, we find that either one
(for zero initial spin) or two compact objects form with masses in
excess of $10^{6}M_{\odot}$ and radii $< 1$~pc (see Fig.~5). In the
case of nonzero spin a binary system of clumps has formed with a
separation of $\sim 10$~pc. Such a system of two compact objects is
expected to efficiently radiate gravitational waves that could be
detectable with the planned {\it Laser Interferometer Space
Antenna}\footnote{See http://lisa.jpl.nasa.gov/} (LISA)\cite{WyL2003}.

What is the further fate of the central object? Once the gas has
collapsed to densities above $\sim 10^{17}$~cm$^{-3}$ and radii
$<10^{16}$~cm, Thomson scattering traps the photons, and the cooling
time consequently becomes much larger than both the free-fall and
viscous timescales (see \cite{BrL2003} for details). The gas is
therefore likely to settle into a radiation-pressure supported
configuration resembling a rotating supermassive star. Recent
fully-relativistic calculations of the evolution of such stars predict
that they would inevitably collapse to a massive black hole
\cite{BS1999}. Under a wide range of initial conditions, a
substantial fraction ($\sim 90$\%) of the mass of the supermassive
star is expected to end up in the black hole.

Is such a complete destruction of H$_{2}$ possible? When we include an
external background of soft UV radiation in our simulation, we find
that a flux level comparable to what is expected close to the end of
the reionization epoch is sufficient to suppress H$_2$ molecule
formation. This is the case even when the effect of self-shielding is
taken into account.  The effective suppression of H$_{2}$
formation crucially depends on the presence of a stellar-like
radiation background. It is therefore likely that stars existed before
the first quasars could have formed.

\begin{theacknowledgments}
VB thanks the Institute for Advanced Study for its hospitality during
part of the work on this review, and expresses his gratitude to Paolo
Coppi and Richard Larson for the many discussions on the first stars.
AL acknowledges support from the Institute for Advanced Study at
Princeton, the John Simon Guggenheim Memorial Fellowship, and NSF
grants AST-0071019, AST-0204514.

\end{theacknowledgments}

\bibliographystyle{aipproc}

\begin{thebibliography}
\expandafter\ifx\csname natexlab\endcsname\relax\def\natexlab#1{#1}\fi
\providecommand{\enquote}[1]{``#1''}
\expandafter\ifx\csname url\endcsname\relax
  \def\url#1{\texttt{#1}}\fi
\expandafter\ifx\csname urlprefix\endcsname\relax\def\urlprefix{URL }\fi

\bibitem[1]{WyL03a}Wyithe, J. S. B., \& Loeb, A. 2003a, ApJ, in press
(astro-ph/0209056)
\bibitem[2]{Cen03}Cen, R. 2003, ApJ, submitted
(astro-ph/0210473)
\bibitem[3]{FL03}Furlanetto, S. R., \& Loeb, A. 2003, ApJ, in press
(astro-ph/0211496)
\bibitem[4]{BL2001}Barkana, R., \& Loeb, A. 2001, Physics Reports, 349, 125
\bibitem[5]{Kap02}Kaplinghat, M., Chu, M., Haiman, Z.,
Holder, G., Knox, L., \& Skordis, C. 2002, ApJ, submitted (astro-ph/0207591)
\bibitem[6]{Yos2003}Yoshida, N., Abel, T., Hernquist, L., \& Sugiyama, N.
2003, ApJ, submitted
\bibitem[7]{HL2001}Haiman, Z., \& Loeb, A. 2001, ApJ, 552, 459
\bibitem[8]{Fan2003}Fan, X., et al. 2003, AJ, in press
(astro-ph/0301135)
\bibitem[9]{BCL1999}Bromm, V., Coppi, P. S., \& Larson, R. B. 1999, ApJ, 527, L5
\bibitem[10]{BCL2002}Bromm, V., Coppi, P. S., \& Larson, R. B. 2002, ApJ, 564, 23
\bibitem[11]{NaU2001}Nakamura, F., \& Umemura, M. 2001, ApJ, 548, 19
\bibitem[12]{ABN2002}Abel, T., Bryan, G. L., \& Norman, M. L. 2002, Science, 295, 93
\bibitem[13]{Pud2002}Pudritz, R. E. 2002, Science, 295, 68
\bibitem[14]{Lar2002}Larson, R. B. 2002, MNRAS, 332, 155
\bibitem[15]{Kr02}Kroupa, P. 2002, Science, 295, 82
\bibitem[16]{MJR1976}Rees, M. J. 1976, MNRAS, 176, 483
\bibitem[17]{BBB2002}Bate, M. R., Bonnell, I. A., \& Bromm, V. 2002, MNARS, 332, L65
\bibitem[18]{BBB2003}Bate, M. R., Bonnell, I. A., \& Bromm, V. 2003, MNRAS, 
in press (astro-ph/0212380)


\bibitem[19]{BrL2003}Bromm, V., \& Loeb, A. 2003, ApJ, submitted 
(astro-ph/0212400)
\bibitem[20]{BKL2001}Bromm, V., Kudritzki, R. P., \& Loeb, A. 2001, ApJ, 552, 464
\bibitem[21]{Cr02}Christlieb, N., et al. 2002, Nature, 419, 904
\bibitem[22]{MBH03}Mackey, J., Bromm, V., \& Hernquist, L. 2003, ApJ, in press
(astro-ph/0208447)


\bibitem[23]{ON1998}Omukai, K., \& Nishi, R. 1998, ApJ, 508, 141
\bibitem[24]{OP2001}Omukai, K., \& Palla, F. 2001, ApJ, 561, L55
\bibitem[25]{Rip2002}Ripamonti, E., Haardt, F., Ferrara, A., \& Colpi, M.
2002, MNRAS, 334, 401
\bibitem[26]{Tan2003}Tan, J. C., \& McKee, C. F. 2003, these proceedings
\bibitem[27]{PSS1983}Palla, F., Salpeter, E. E., \& Stahler, S. W. 1983, ApJ, 271, 632
\bibitem[28]{OI2002}Omukai, K., \& Inutsuka, S. 2002, MNRAS, 332, 59
\bibitem[29]{WC1987}Wolfire, M. G., \& Cassinelli, J. P. 1987, ApJ, 319, 850
\bibitem[30]{MFR01}Madau, P., Ferrara, A., \& Rees, M. J. 2001, ApJ, 555, 92
\bibitem[31]{MFM02}Mori, M., Ferrara, A., \& Madau, P. 2002, ApJ, 571, 40
\bibitem[32]{TSD02}Thacker, R.J., Scannapieco, E., \& Davis, M. 2002, ApJ, 581, 836
\bibitem[33]{LH97} Loeb, A.,\& Haiman, Z.\ 
1997, ApJ, 490, 571
\bibitem[34]{TodF01} Todini, P., \& Ferrara, A. 2001, MNRAS, 325, 726
\bibitem[35]{MBA01}Machacek, M. E., Bryan, G. L., \& Abel, T. 2001, ApJ, 548, 509
\bibitem[36]{MBA03}Machacek, M. E., Bryan, G. L., \& Abel, T. 2003, MNRAS, 338, 27
\bibitem[37]{Om00}Omukai, K. 2000, ApJ, 534, 809
\bibitem[38]{BFCL01}Bromm, V., Ferrara, A., Coppi, P. S., \& Larson, R. B. 2001, MNRAS, 328, 969
\bibitem[39]{Sch02} Schneider, R., Ferrara, A., Natarajan, 
P., \& Omukai, K.\ 2002, ApJ, 571, 30 
\bibitem[40]{LR00}Lamb, D. Q., \& Reichart, D. E. 2000, ApJ, 536, 1
\bibitem[41]{Cia00} Ciardi, B., \& Loeb, A.\ 2000, ApJ, 540, 687 
\bibitem[42]{Bloom01} Bloom, 
J.~S., Kulkarni, S.~R., \& Djorgovski, S.~G.\ 2002, AJ, 123, 1111
\bibitem[43]{kul} Kulkarni, S. R., et 
al.\ 2000, Proc. SPIE, 4005, 9  
\bibitem[44]{Tot97}Totani, T. 1997, ApJ, 486, L71
\bibitem[45]{Wij98}Wijers, R. A. M. J., Bloom, J. S., Bagla, J. S., 
\& Natarajan, P. 1998, MNRAS, 294, L13
\bibitem[46]{BlNat00}Blain, A. W., \& Natarajan, P. 2000, MNRAS, 312, L35
\bibitem[47]{BrL2002}Bromm, V., \& Loeb, A. 2002, ApJ, 575, 111

\bibitem[48]{Heg03} Heger, A., Fryer, C. L., Woosley, S. E.,
Langer, N., \& Hartmann, D. H. 2003, ApJ, submitted
(astro-ph/0212469)
\bibitem[49]{BL03} Barkana, R., \& Loeb, A. 2003, Nature, in press
(astro-ph/0209515)
\bibitem[50]{Mi98} Miralda-Escud\'{e}, J.\ 
1998, ApJ, 501, 15 
\bibitem[51]{Rees1984}Rees, M. J. 1984, ARA\&A, 22, 471
\bibitem[52]{Beck2001}Becker, R. H., et al. 2001, AJ, 122, 2850
\bibitem[53]{Djor2001}Djorgovski, S.G., Castro, S., Stern, D.,
\& Mahabal, A. A. 2001, ApJ, 560, L5
\bibitem[54]{Fan2002}Fan, X., Narayanan, V. K., Strauss, M. A., White, R. L.,
Becker, R. H., Pentericci, L., \& Rix, H.-W.  2002, AJ, 123, 1247
\bibitem[55]{LP02} Loeb, A., \& Peebles, P. J. E. 2002, ApJ, submitted
(astro-ph/0211465)
\bibitem[56]{LR1994}Loeb, A., \& Rasio, F. A. 1994, ApJ, 432, 52
\bibitem[57]{OH02}Oh, S.P., \& Haiman, Z. 2002, ApJ, 569, 558
\bibitem[58]{HRL1997}Haiman, Z., Rees, M. J., \& Loeb, A. 1997, ApJ, 476, 458;
erratum 484, 985
\bibitem[59]{WyL2003}Wyithe, J. S. B., \& Loeb, A. 2003b, ApJ, submitted 
(astro-ph/0211556)
\bibitem[60]{BS1999} Baumgarte, T. W., \& Shapiro, S. L. 1999, ApJ, 526, 941

\end{thebibliography}

\end{document}